\documentclass[
superscriptaddress,
aps,prl,
reprint,
twocolumn,
amsmath,amssymb,showpacs
]{revtex4-2}
\usepackage{newtxtext,newtxmath}
\usepackage{physics}
\usepackage{graphicx}
\usepackage{microtype}
\usepackage{dsfont}
\usepackage{xcolor}
\usepackage{siunitx}
\usepackage{multirow}
\definecolor{urlc}{RGB}{58,105,157}
\usepackage[
colorlinks=true,
urlcolor=urlc,
linkcolor=urlc,
citecolor=urlc
]{hyperref}

\usepackage{colortbl}
\definecolor{color1}{RGB}{239,247,247}

\begin{document}
\widetext

\title{High-Accuracy Temporal Prediction via Experimental Quantum Reservoir Computing in Correlated Spins}

\author{Yanjun Hou}
\thanks{These authors contribute equally}
\affiliation{
Laboratory of Spin Magnetic Resonance, School of Physical Sciences, Anhui Province Key Laboratory of Scientific Instrument Development and Application, University of Science and Technology of China, Hefei 230026, China}

\author{Juncheng Hua}
\thanks{These authors contribute equally}
\affiliation{
Laboratory of Spin Magnetic Resonance, School of Physical Sciences, Anhui Province Key Laboratory of Scientific Instrument Development and Application, University of Science and Technology of China, Hefei 230026, China}
\affiliation{
Hefei National Research Center for Physical Sciences at the Microscale, University of Science and Technology of China, Hefei 230026, China}

\author{Ze Wu}
\affiliation{
Laboratory of Spin Magnetic Resonance, School of Physical Sciences, Anhui Province Key Laboratory of Scientific Instrument Development and Application, University of Science and Technology of China, Hefei 230026, China}
\affiliation{
Department of Physics and The Hong Kong Institute of Quantum Information Science and Technology,
The Chinese University of Hong Kong, Shatin, New Territories, Hong Kong, China}

\author{Wei Xia}
\affiliation{
Department of Physics and The Hong Kong Institute of Quantum Information Science and Technology,
The Chinese University of Hong Kong, Shatin, New Territories, Hong Kong, China}

\author{Yuquan Chen}
\affiliation{
Laboratory of Spin Magnetic Resonance, School of Physical Sciences, Anhui Province Key Laboratory of Scientific Instrument Development and Application, University of Science and Technology of China, Hefei 230026, China}

\author{Xiaopeng Li}
\email{xiaopeng\underline{ }li@fudan.edu.cn}
\affiliation{State Key Laboratory of Surface Physics, Institute of Nanoelectronics and Quantum Computing, and Department of Physics, Fudan University, Shanghai 200433, China}
\affiliation{Shanghai Qi Zhi Institute, and Shanghai Artificial Intelligence Laboratory, Xuhui District, Shanghai 200232, China} 
\affiliation{Shanghai Research Center for Quantum Sciences, Shanghai 201315, China}
\affiliation{
Hefei National Laboratory, University of Science and Technology of China, Hefei 230088, China}

\author{Zhaokai Li}
\email{zkli@ustc.edu.cn}
\affiliation{
Laboratory of Spin Magnetic Resonance, School of Physical Sciences, Anhui Province Key Laboratory of Scientific Instrument Development and Application, University of Science and Technology of China, Hefei 230026, China}
\affiliation{
Hefei National Research Center for Physical Sciences at the Microscale, University of Science and Technology of China, Hefei 230026, China}
\affiliation{
Hefei National Laboratory, University of Science and Technology of China, Hefei 230088, China}

\author{Xinhua Peng}
\email{xhpeng@ustc.edu.cn}
\affiliation{
Laboratory of Spin Magnetic Resonance, School of Physical Sciences, Anhui Province Key Laboratory of Scientific Instrument Development and Application, University of Science and Technology of China, Hefei 230026, China}
\affiliation{
Hefei National Research Center for Physical Sciences at the Microscale, University of Science and Technology of China, Hefei 230026, China}
\affiliation{
Hefei National Laboratory, University of Science and Technology of China, Hefei 230088, China}

\author{Jiangfeng Du}
\email{djf@ustc.edu.cn}
\affiliation{
State Key Laboratory of Ocean Sensing and School of Physics, Zhejiang University, Hangzhou 310058, China}

\begin{abstract}
Physical reservoir computing provides a powerful machine learning paradigm that exploits nonlinear physical dynamics for efficient information processing. 
By incorporating quantum effects, quantum reservoir computing offers superior potential for machine learning applications, as quantum dynamics are exponentially costly to simulate classically. 
Here, we present a novel quantum reservoir computing approach based on correlated quantum spin systems, exploiting natural quantum many-body interactions to generate reservoir dynamics, thereby circumventing the practical challenges of deep quantum circuits.
Our experimental implementation supports nontrivial quantum entanglement and exhibits sufficient dynamical complexity for high-performance machine learning.
We achieve state-of-the-art performance in experiments on standard time-series benchmarks, reducing prediction error by 1 to 2 orders of magnitude compared to previous quantum reservoir experiments.
In long-term weather forecasting, our 9-spin quantum reservoir delivers greater prediction accuracy than classical reservoirs with thousands of nodes. 
This represents the first experimental demonstration of quantum machine learning outperforming large-scale classical models on real-world tasks. 
\end{abstract}

\maketitle

\emph{Introduction}---The rapid growth of artificial intelligence has highlighted the demand for novel computational paradigms capable of real-time processing of temporally structured, data-intensive information streams.
Physical reservoir computing (PRC) offers a promising solution by harnessing the intrinsic dynamics of physical substrates as computational resources within a machine learning framework \cite{tanaka2019recent, nakajima2020physical}. 
In this approach, inputs are projected into the high-dimensional space of a fixed dynamical system that simultaneously performs processing and memory, thus avoiding the von Neumann bottleneck. 
The output is obtained by only training a linear readout layer, enabling fast inference with minimal learning cost.
PRC can be implemented in a broad range of physical systems that provide a high-dimensional state space, nonlinear input–output mapping, and fading memory \cite{konkoli2017reservoir}, including memristors \cite{moon2019temporal, milano2022materia, liu2022multilayer, liang2024physical}, spintronics \cite{torrejon2017neuromorphic, romera2018vowel, grollier2020neuromorphic, lee2024task}, and optical systems \cite{brunner2013parallel, vandoorne2014experimental, larger2017high, rafayelyan2020large}.
As larger and more complex dynamical systems tend to enhance computational capacity \cite{kaplan2020scaling, raghu2017expressive, lukovsevivcius2009reservoir}, this framework naturally extends into the quantum regime. 
Quantum reservoir computing (QRC) leverages the exponentially large Hilbert space and complex correlations of quantum systems to offer unique advantages in solving challenging tasks. 
Given that some noisy intermediate-scale quantum (NISQ) systems have already outperformed classical simulators \cite{arute2019quantum, zhong2020quantum, madsen2022quantum}, current quantum devices hold strong potential for realizing QRC with computational capacity and expressiveness beyond classical models.

Despite its strong potential, QRC has so far been explored primarily through theoretical studies and numerical simulations on various idealized quantum models, while experimental realizations remain relatively limited \cite{mujal2021opportunities, ghosh2021quantum}.
Most implementations rely on deep quantum circuits to generate the complex dynamics required for QRC \cite{hu2024overcoming, pfeffer2022hybrid, chen2020temporal, kubota2023temporal, suzuki2022natural, yasuda2023quantum, monzani2024leveraging}.
However, such circuits often demand substantial depth, making them vulnerable to gate errors and dissipation, thereby limiting the fidelity and performance of the QRC in the experiments.
Furthermore, while quantum systems inherently possess exponentially large Hilbert spaces, current readout schemes are typically limited to local measurements, restricting the full exploitation of the high-dimensional nature of the quantum dynamics.

In this Letter, we present an experimental realization of quantum reservoir computing based on a quantum spin system. 
To address the abovementioned challenges, we introduce two key innovations: the integration of both coherent spin-spin interactions and incoherent spin relaxation processes to enrich the dynamical complexity of the reservoir, and the use of time-multiplexed measurements to increase the effective readout capacity of the reservoir \cite{fujii2017harnessing}. 
Building on these, we achieve the state-of-the-art performance in QRC experiments on the widely used NARMA benchmark task, reducing the prediction error by 1 to 2 orders of magnitude compared to previous results \cite{suzuki2022natural, yasuda2023quantum, monzani2024leveraging}. 
We apply the QRC protocol to weather forecasting, which is crucial for countless social and economic activities \cite{price2025probabilistic}, yet challenging due to the chaotic nature of the atmosphere \cite{kalnay2003atmospheric, palmer1993extended}.
In long-term predictions, our 9-spin quantum reservoir achieves higher prediction accuracy than classical reservoirs with thousands of nodes. 
Our results suggest practical quantum advantages in time-series prediction may be achievable with current quantum hardware.
To our knowledge, this represents a first experimental demonstration of a quantum machine learning system that outperforms large-scale classical networks on realistic datasets, where previous advantages were typically shown via numerical simulations or limited to synthetic datasets \cite{biamonte2017quantum, huang2021power, liu2021rigorous, cerezo2022challenges, caro2022generalization, jager2023universal, xia2023configured, kornjavca2024large}.

\begin{figure}
    \centering
    \includegraphics[scale=1]{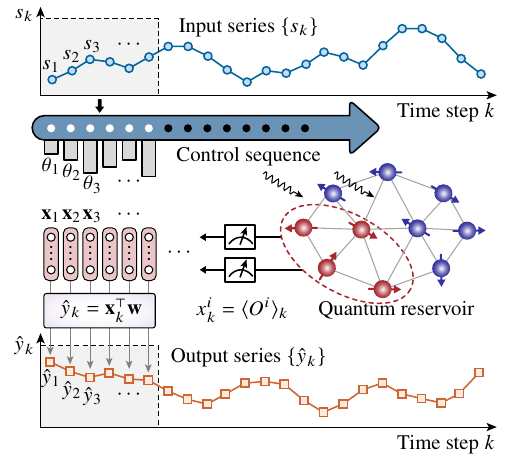}
    \caption{
    Schematic diagram of quantum reservoir computing.
    The input series $\{s_k\}$ is mapped to the control parameters $\{\theta_k\}$ of a sequence of quantum operations, which are sequentially applied to a quantum reservoir comprising a spin network. 
    The reservoir state is probed through measurements of partial observables, producing a readout vector with components $x^{i}_k = \langle O^i \rangle_k = {\rm Tr}(\rho_k O^i)$. 
    The final output is computed as a linear combination, $\hat{y}_k={\bf x}_{k}^{\top} {\bf w}$. 
    }
    \label{fig:1}
\end{figure}

\emph{Model design}---We introduce the general framework of QRC for temporal information processing, as illustrated in Fig.~\ref{fig:1}.
Consider a pair of discrete scalar sequences: the input $\{s_{k}\}$ and the target output $\{y_{k}\}$, where $y_k$ depends on a finite input history up to time step $k$, denoted by $y_{k} = f(s_{k}, \ldots, s_{k-\Delta})$.
The goal of QRC is to approximate this relationship by generating an output sequence $\{\hat{y}_{k}\}$ that closely matches the target.
The inputs $\{s_{k}\}$ are encoded into a control sequence and sequentially applied to an $N$-qubit quantum reservoir, mapping the information into a $4^N$-dimensional Hilbert space.
Between successive inputs, the reservoir undergoes a fixed intrinsic evolution, which integrates new inputs with the memory of past inputs.
To extract the information, specific observables are measured at each time step, yielding a sequence of readout vectors ${{\bf x}_k}$ with $x_k^i = {\rm Tr}(\rho_k O^i)$.
The output is computed as a weighted sum of readout components, $\hat{y}_k = {\bf x}_k^\top {\bf w}$, with a bias term incorporated by augmenting ${\bf x}_k$ with $x^0_k = 1$.
Following standard machine learning practices, the readout weights ${\bf w}$ are optimized on a training dataset by minimizing the error $\sum_{k=1}^{L} \left(y_k - \hat{y}_k \right)^2$, and the performance of the resulting QRC is then evaluated on test data.

\begin{figure*}
    \centering
    \includegraphics[scale=1]{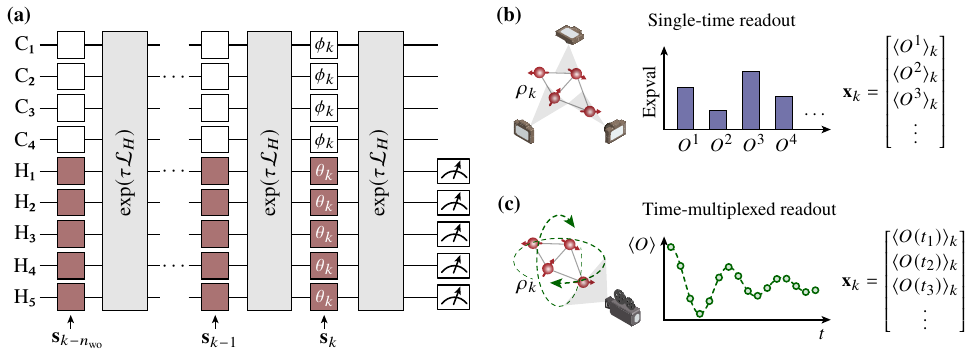}
    \caption{
    Schematic of experimental setup and readout schemes.
    {\bf (a)} Circuit diagram of the QRC experimental implementation. 
    The reservoir, composed of carbon and proton spins, undergoes repeated cycles of input-dependent rotations $\{\phi_k,\theta_k\}$ followed by intrinsic evolution under $\exp(\tau\mathcal{L}_H)$.
    The experiment adopts a rewinding protocol, where each experiment at time step $k$ is restarted from cycle $k-n_{\rm wo}$ to wash out initial state dependence, and readout is performed on the proton spins at the end of the circuit.
    Two readout schemes are considered.
    {\bf (b)} Single-time readout: analogous to capturing multiple images of a static system by artificially changing the viewpoint.
    A fixed set of observables $\{O^i\}$ (e.g. Pauli operators) is measured on the reservoir state $\rho_k$, yielding the readout vector ${\bf x}_k = [\langle O^1\rangle_k, \langle O^2\rangle_k, \dots]^\top$ where $\langle O\rangle_k={\rm Tr}(\rho_k O)$.
    {\bf (c)} Time-multiplexed readout: analogous to filming a moving system from a fixed viewpoint, where other aspects naturally rotate into view.
    A single observable $O$ is sampled at multiple time points during the reservoir's evolution, equivalently measuring the Heisenberg‑picture operators $\{O(t_i)\}$, yielding the readout vector ${\bf x}_k = [\langle O(t_1)\rangle_k, \langle O(t_2)\rangle_k, \dots]^\top$.
    }
    \label{fig:2}
\end{figure*}

The computational capabilities of QRC essentially depend on both the dynamics of the quantum reservoir and the number of readout components \cite{martinez2021dynamical, xia2022reservoir, dambre2012information, martinez2023information}.
Generally, QRC dynamics consist of unitary and nonunitary components.
The unitary evolution, realized via analog Hamiltonian dynamics or digital quantum circuits, transforms and spreads information throughout the system while preserving coherence and generating entanglement. 
In addition to enriching the dynamics, the nonunitary part is essential for retaining memory of recent inputs by gradually diminishing the influence of past states, a property known as fading memory \cite{boyd1985fading}. 
Previous protocols have achieved this through qubit resetting \cite{fujii2017harnessing, hu2024overcoming} or engineered dissipative channels \cite{chen2020temporal, sannia2024dissipation}, but these methods often require intricate control techniques or additional experimental resources.

Here, we adopt a more efficient method by leveraging the inherent relaxation which commonly exists in natural quantum systems.
Thus, the system evolution is governed by the Lindbladian superoperator $\dot{\rho} = \mathcal{L}_{H} [\rho] = -i \left[H, \rho \right] + \mathcal{R}[\rho]$, where $H$ is the system's Hamiltonian and $\mathcal{R}$ denotes the relaxation channel.
The overall state update is then expressed as
\begin{equation}
    \rho_{k} = e^{\tau \mathcal{L}_{H}} \circ \mathcal{U}_{s_k} \left[\rho_{k-1} \right], 
    \label{eq:state}
\end{equation}
where $e^{\tau \mathcal{L}_{H}}$ denotes the propagator for evolution over duration $\tau$, and $\mathcal{U}_{s_k} [\rho] = U_{s_k} \rho U^{\dagger}_{s_k}$ represents the unitary channel used to encode the classical input $s_k$.
In spin systems, two dominant relaxation processes are longitudinal relaxation (with time constant $T_1$), which drives the system toward thermal equilibrium via energy dissipation, and transverse relaxation (with time constant $T_2$), which causes coherence decay through phase randomization \cite{levitt2008spin}.
These processes correspond to generalized amplitude damping and dephasing channels, respectively \cite{nielsen2001quantum}.
Our results indicate that $T_1$ relaxation plays a pivotal role in ensuring QRC performance (see Supplemental Material \cite{supp}), in agreement with recent numerical studies \cite{domingo2023taking, kubota2023temporal, monzani2024leveraging}.
By incorporating inevitable relaxation as a computational resource, our method overcomes coherence time limitations and eliminates the need for qubit resets, enabling prolonged experimental temporal processing with minimal control.

Despite the rich dynamics of quantum reservoirs, the information processing capacity of a dynamical system is fundamentally limited by the number of independent readout functions \cite{dambre2012information}.
While quantum systems provide access to an exponentially large state space, extracting complete information through state tomography becomes increasingly challenging for large systems, particularly when high-order correlations are involved \cite{nielsen2001quantum}.
Consequently, most existing schemes rely on measuring local observables like $\langle \sigma_z^i \rangle$ \cite{chen2020temporal, suzuki2022natural, kubota2023temporal, yasuda2023quantum, monzani2024leveraging}, which capture only a small subset of the reservoir's degrees of freedom, thus failing to fully exploit its computational potential.

To address this limitation, we employ time multiplexing, a technique that effectively increases the number of readouts by measuring fixed observables at different times $\{t_i\}$ during the reservoir's evolution \cite{fujii2017harnessing}, as illustrated in Fig.~\ref{fig:2}(c). 
By accessing the Heisenberg-evolved observables $\{O(t_i)\}$, this scheme provides more independent information about the reservoir state than a single-time measurement can offer, thereby raising the upper bound on the achievable performance \cite{supp}.
In our implementation, we utilize the free induction decay (FID) signal, a natural time-multiplexed readout in quantum systems, with experimental results showing significant performance improvements.

\emph{High-accuracy temporal prediction}---To demonstrate the performance of our approach, we perform experimental benchmarks on a nuclear magnetic resonance (NMR) quantum platform.
The initial task we consider is the nonlinear autoregressive moving average (NARMA) simulation, a widely adopted benchmark for temporal processing capabilities \cite{hochreiter1997long, atiya2000new}.
In accordance with previous work \cite{fujii2017harnessing,suzuki2022natural, yasuda2023quantum, monzani2024leveraging}, the input sequence is constructed as a superposition of sine waves, while the target sequence is generated by specific recurrent dynamics, for example:
\begin{equation}
    y_{k+1}=\alpha y_k+\beta y_k\left(\sum_{i=0}^{n-1} y_{k-i}\right)+\gamma s_{k-n+1} s_k+\delta,
\end{equation}
where $n$ determines the number of past time steps influencing the current output, with the task denoted by NARMA$n$.
We allocate 400 steps for training and 100 steps for testing.
Readout weights are optimized via ridge regression: ${\bf w}^{*}=\left({\bf X}^{\top} {\bf X} + \lambda {\bf I} \right)^{-1} {\bf X}^{\top} {\bf y}$, where ${\bf X}=\left[ {\bf x}_1, \ldots, {\bf x}_{400} \right]^{\top}$ and ${\bf y}=\left[ y_1, \ldots, y_{400} \right]^{\top}$ denote the readout matrix and target vector for training, and $\lambda$ is the regularization strength \cite{hoerl1970ridge}.
By fitting separate ${\bf w}^*$ for each $n$ on the same ${\bf X}$, the reservoir simultaneously emulates multiple NARMA systems, a capability referred to as multitasking \cite{fujii2017harnessing}.

Figure~\ref{fig:2}(a) presents the circuit schematic for our experiments.
The quantum reservoir is a 9-spin coupling system consisting of four carbon nuclei and five proton nuclei in a $^{13}$C-labeled crotonic acid molecule.
The spin reservoir undergoes successive cycles of input encoding and free evolution according to Eq.~\eqref{eq:state}.
For NARMA task, its scalar input is normalized to $\bar{s}_k\in [0,1]$ and encoded via an $x$-axis radio-frequency pulse on proton nuclei, effecting a global rotation $U_{s_k}=\prod_{n=5}^9 R^{n}_x\left[\arcsin(\bar{s}_k)\right]$, where $R^{n}_x(\theta)=e^{-i\frac{\theta}{2}\sigma^n_x}$ and $\sigma^n_{x,y,z}$ are Pauli operators on spin $n$.
The system Hamiltonian is 
\begin{equation}
    H=\sum_{i=1}^{9} \pi \nu_i \sigma^{i}_{z} + \sum_{1\leq i<j \leq 9} \frac{\pi}{2} J_{ij} \sigma^{i}_{z} \sigma^{j}_{z},
    \label{eq:Hamiltonian}
\end{equation}
where $\nu_i$ represents the chemical shift of spin $i$ and $J_{ij}$ denotes the coupling between spin $i$ and spin $j$ \cite{supp}.
To circumvent the measurement backaction, we adopt the rewinding protocol \cite{mujal2023time}, where the experiment of time step $k$ is performed with input cycles from $k-n_{\rm wo}$ to $k$.
The first $n_{\rm wo}$ cycles serve to wash out the initial state dependence under the premise of fading memory, thus allowing the experiment to start from any state without specific preparation.
The washout length $n_{\rm wo}$ depends on the system's relaxation rates and evolution durations, and is determined through preliminary experiments \cite{supp}.

After the $n_{\rm wo}+1$ cycles, readout is performed on the proton spins using two schemes, as illustrated in Figs.~\ref{fig:2}(b) and (c).
Single‑time readout refers to the conventional measurement of a fixed set of observables, here chosen as single-qubit Pauli operators. 
As $\rm H_3$–$\rm H_5$ are indistinguishable methyl protons, only nine distinct expectation values, $\langle \sigma^{m}_{\alpha} \rangle$ ($\alpha=x,y,z$, $m={\rm H_1}, {\rm H_2}, {\rm methyl}$), are accessible.
Time-multiplexed readout is implemented through the standard NMR FID detection, which monitors the transverse magnetization via continuous weak measurement~\cite{levitt2008spin}, requiring no extra experimental overhead~\cite{supp}. 
The recorded signal can be written as
\begin{equation}
    S(t) \propto {\rm Tr}\left\{e^{t\mathcal{L}_H} \left[U \rho U^\dagger \right] O_{\rm FID}\right\},
\end{equation}
where $O_{\rm FID}=\sum_{m=5}^{9} \left(\sigma^m_y+i \sigma^m_x\right)$ is the collective transverse operator on protons.
In the experiments, we apply a $\pi/2$ pulse before readout ($U=\prod_{n=5}^9 R^n_x(\pi/2)$), and perform a Fourier transform on the FID signal to extract the reservoir response, selecting spectral peaks to form 653 readout features.
Further details are provided in Supplemental Material \cite{supp}.

\begin{figure}
    \centering
    \includegraphics[scale=1]{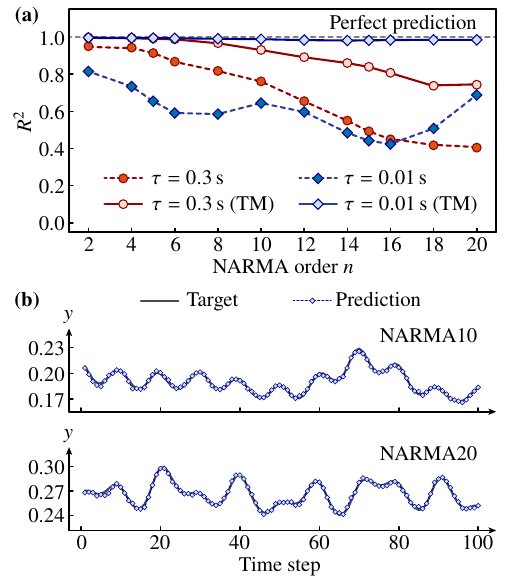}
    \caption{
    \textbf{High accuracy temporal prediction with NARMA benchmark.}
    {\bf (a)} Prediction performance (evaluated by $R^2$) on NARMA tasks with $n$ varying from 2 to 20, using quantum reservoirs with different configurations.
    Red and blue lines represent evolution durations of $\tau = \SI{0.3}{s}$ and $\SI{0.01}{s}$, respectively, while solid and dashed lines indicate results obtained with time-multiplexed (TM) and single-time readout schemes.
    {\bf (b)} Prediction results for NARMA tasks with $n=10$ and $20$, utilizing time-multiplexed readout with $\tau=\SI{0.01}{s}$.
    }
    \label{fig:3}
\end{figure}

\begin{table}
\setlength\tabcolsep{0pt}
\renewcommand{\arraystretch}{1.3}
\centering
\small
\begin{tabular}{|p{1.8cm}<{\centering}|p{1.7cm}<{\centering}|p{1.7cm}<{\centering}|p{1.7cm}<{\centering}|p{1.7cm}<{\centering}|}
    \hline
    \cellcolor{color1} & \multicolumn{2}{c|}{\cellcolor{color1}$\tau=\SI{0.3}{s}$} & \multicolumn{2}{c|}{\cellcolor{color1}$\tau=\SI{0.01}{s}$}  \\ 
    \cline{2-5} 
    \multirow{-2}{*}{\cellcolor{color1}NMSE} & \cellcolor{color1} & \cellcolor{color1}TM & \cellcolor{color1} & \cellcolor{color1}TM \\ 
    \hline
    \cellcolor{color1}NARMA2  & $2.76 \times 10^{-6}$ & $\bf 1.74 \times 10^{-7}$ & $1.03 \times 10^{-5}$ & $2.10 \times 10^{-7}$ \\ \hline
    \cellcolor{color1}NARMA5  & $5.03 \times 10^{-4}$ & $4.44 \times 10^{-5}$ & $2.01 \times 10^{-3}$ & $\bf 3.44 \times 10^{-5}$ \\ \hline
    \cellcolor{color1}NARMA10 & $1.16 \times 10^{-3}$ & $3.39 \times 10^{-4}$ & $1.73 \times 10^{-3}$ & $\bf 5.84 \times 10^{-5}$ \\ \hline
    \cellcolor{color1}NARMA15 & $1.78 \times 10^{-3}$ & $5.64 \times 10^{-4}$ & $1.96 \times 10^{-3}$ & $\bf 6.37 \times 10^{-5}$ \\ \hline
    \cellcolor{color1}NARMA20 & $1.66 \times 10^{-3}$ & $7.15 \times 10^{-4}$ & $8.68 \times 10^{-4}$ & $\bf 4.34 \times 10^{-5}$ \\ \hline
\end{tabular}
\caption{
    List of normalized mean squared errors ($\mathrm{NMSE} = \sum_{k=1}^{L} \left( y_k - \hat{y}_k \right)^2 \big/ \sum_{k=1}^{L} y_k^2$) for NARMA tasks with orders $n = 2$, 5, 10, 15, and 20, using QRC under various configurations.
    The minimum error in each task is highlighted.
    }
\label{tab:nmse}
\end{table}

We evaluate quantum reservoirs with different evolution durations and readout schemes on NARMA tasks of orders $n=2$ to 20. 
The results are shown in Fig.~\ref{fig:3}(a), with performance quantified by the coefficient of determination,
\begin{equation}
    R^2=1-\frac{\sum_{k=1}^{L}\left( y_k - \hat{y}_k \right)^2}{\sum_{k=1}^{L}\left( y_k - \bar{y} \right)^2},
\end{equation}
where $\bar{y}=\frac{1}{L}\sum_{k=1}^{L} y_k$.
For direct comparison with prior work, we also report the corresponding normalized mean square errors in Table~\ref{tab:nmse}; our best cases achieve error reductions of 1-2 orders of magnitude relative to recent experimental benchmarks \cite{suzuki2022natural, yasuda2023quantum, monzani2024leveraging}.
Time-multiplexed readout markedly improves performance over single-time readout, especially for $\tau = \SI{0.01}{s}$, where predictions are nearly perfect across all NARMA orders (see NARMA10 and 20 in Fig.~\ref{fig:3}(b) and additional results in Supplemental Material \cite{supp}).
As the relaxation effects become more pronounced with longer evolution times, the information from high-order correlations is significantly diminished at $\tau=\SI{0.3}{s}$, thereby reducing the benefit of time multiplexing. 

In Supplemental Material~\cite{supp}, we show that the quantum reservoir markedly outperforms its classical spin-based counterpart, which yields an average $R^2 < 0.75$ across NARMA2 to NARMA20, highlighting the computational role of quantum correlations. 
We also provide a quantitative assessment of the protocol's robustness against several experimental imperfections, offering guidance for implementing the scheme on various physical platforms.

\begin{figure}
    \centering
    \includegraphics[scale=1]{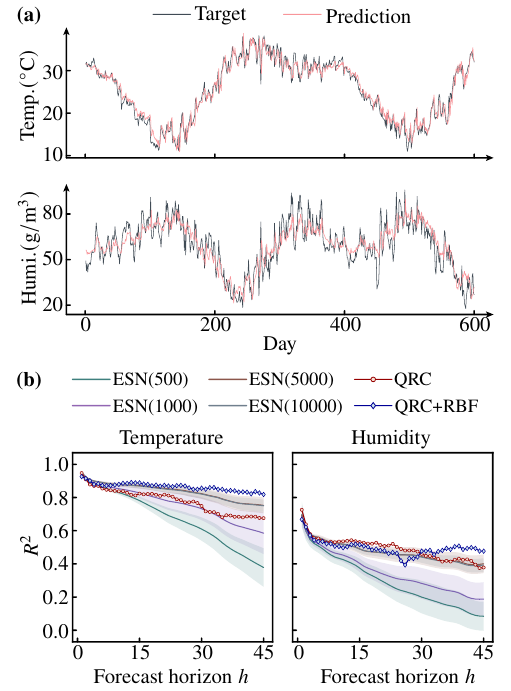}
    \caption{
    Weather forecasting via quantum reservoir computing.
    {\bf (a)} Experimental results for single-step-ahead forecasting of temperature and humidity using quantum reservoir computing. 
    The black and red lines denote the target and predicted trajectories, respectively.
    {\bf (b)} Multi-step-ahead forecasting performance comparison between echo state networks (ESNs) with varying node counts and QRC methods. 
    $R^2$ is averaged over 100 random realizations for each ESN($m$), with shaded regions representing the standard deviation as error bars.
    }
    \label{fig:4}
\end{figure}

\emph{Realistic weather forecasting}---To show the practical utility of our approach, we conduct weather forecasting on a daily climate dataset from Delhi, a benchmark commonly employed in the machine learning community \cite{daily-climate}.
Our analysis focuses on predicting the multivariate time series of temperature and humidity, ${\bf y}_k = \left[{\rm Temp}(k), {\rm Humi}(k)\right]$.
We first perform single-step-ahead forecasting, where ${\bf y}_k$ serves as the $k$th input and ${\bf y}_{k+1}$ as the corresponding target. 
The dataset is divided into 374 washout steps, 600 training steps, and 600 testing steps. 
The normalized temperature and humidity data are encoded as angles $\theta_k$ and $\phi_k$ for global rotations applied to the proton and carbon nuclei, respectively, with an input interval of $\tau = \SI{0.03}{s}$.
Figure~\ref{fig:4}(a) presents the prediction results alongside the target values from the test set.
Real weather data exhibit substantial stochastic fluctuations, which pose challenges for precise forecasting. 
Nevertheless, QRC predictions capture the overall trends and closely match local variations in several segments of the time series.

We subsequently assess multi-step-ahead forecasting, in which predictability is constrained by the rapid error growth inherent to the chaotic atmospheric system \cite{kalnay2003atmospheric, palmer1993extended}.
Here, a direct multistep forecast strategy is employed \cite{taieb2012review}, wherein a separate model is trained for each forecast horizon $h$: ${\bf y}_{k+h} = f_{h}({\bf y}_{k}, {\bf y}_{k-1}, {\bf y}_{k-2}, \ldots)$. 
This is implemented through multitasking in QRC, where distinct readout weights are trained for each target sequence $\{{\bf y}_{k+h}\}$ while utilizing the same input sequence $\{{\bf y}_k\}$.
To evaluate the potential quantum advantages, we compare the performance of QRC with that of echo state networks (ESNs) \cite{jaeger2004harnessing}, with results displayed in Fig.~\ref{fig:4}(b).
ESNs with 500, 1000, 5000, and 10000 nodes (denoted as ESN($m$) for $m$-node networks) are evaluated, showing improved performance with increasing size until saturation, where the curves of ESN(5000) and ESN(10000) nearly overlap, indicating diminishing returns.
For temperature forecasting, QRC attains accuracy comparable to that of ESN(1000), and even exceeds its average at larger forecast horizons $h$. 
Moreover, incorporating nonlinear postprocessing may further enhance QRC's performance: support vector regression (SVR) with a radial basis function (RBF) kernel \cite{awad2015support} enables QRC to achieve higher accuracy than ESN(10000), whereas ESNs gain no significant improvement \cite{supp}.
We attribute QRC's strong performance to the superior expressibility \cite{sim2019expressibility} and unique information processing capabilities \cite{kubota2023temporal} inherent to quantum models.
As humidity is strongly influenced by rapidly varying local processes such as convection and precipitation, its prediction becomes inherently more challenging.
Nonetheless, QRC still achieves performance comparable to that of ESN(10000).
These results show the capability of quantum systems to address complex dynamical prediction tasks, highlighting the potential quantum advantages in real-world applications with current NISQ devices.

\emph{Conclusion}---Building on the natural dynamics of a spin system, we develop a hardware-efficient quantum reservoir computing scheme that enables accurate time-series predictions in practical machine learning tasks. 
Implemented on a 9-spin system with time-multiplexed readouts, our approach achieves 1 to 2 orders of magnitude lower error compared to previous QRC experiments on NARMA benchmarks. In long-term weather forecasting, our quantum reservoir achieves higher prediction accuracy than classical reservoirs with thousands of nodes, suggesting that practical quantum advantages in time-series prediction may be attainable with current quantum hardware.
We anticipate that the computational capacity of our QRC protocol can be further improved by increasing the complexity of the reservoir dynamics, for example, through pulse-engineered refinement of internal interactions, or by scaling to larger quantum processors. 
Moreover, integrating QRC with many-body phenomena, including dynamical phase transitions \cite{martinez2021dynamical, xia2022reservoir, kobayashi2023quantum, kobayashi2024quantum}, quantum scar states \cite{serbyn2021quantum, bravo2022quantum}, and time crystals \cite{zhang2017observation}, represents a promising direction for future research, as it could help uncover physical principles to improve QRC performance systematically.  

\emph{Acknowledgments}---This work is supported by Quantum Science and Technology-National Science and Technology Major Project (Grants No.~2024ZD0302000, No.~2021ZD0303205 and No.~2024ZD0300100), National Natural Science Foundation of China (Grants No.~T2388102, No.~12261160569 and No.~92165108), National Basic Research Program of China (Grants No.~2021YFA1400900), Shanghai Municipal Science and Technology (No. 25TQ003, No. 2019SHZDZX01 and No. 24DP2600100), and the XPLORER Prize. The authors acknowledge helpful discussions with the Anhui Meteorological Bureau of China.

\emph{Data availability}---The data and code that support the findings of this study are openly available \cite{Hou2025Zenodo}.

\bibliography{main}

\clearpage
\appendix
\onecolumngrid
\begin{center}
{\bf Supplemental Material for \emph{High-Accuracy Temporal Prediction via Experimental Quantum Reservoir Computing in Correlated Spins}}
\end{center}

\setcounter{secnumdepth}{1}
\section{Experimental setup}

In this section we present the experimental setup employed in this work.
Our experiments were conducted using a Bruker Avance III 400 MHz spectrometer at 298 K. 
The quantum reservoir is a 9-spin coupling system consisting of four carbon nuclei and five proton nuclei in a $^{13}$C-labeled crotonic acid molecule, the structure of which is shown in TABLE~\ref{tab:sample_param}. 
The methyl protons $\rm H_3$, $\rm H_4$, and $\rm H_5$ are chemically and magnetically equivalent and therefore indistinguishable. 
TABLE~\ref{tab:sample_param} lists the chemical shifts $\nu_i$ (diagonal elements) and the $J$-coupling constants $J_{ij}$ (off-diagonal elements), along with the corresponding relaxation times. 
In particular, the $T_2$ times account for dephasing induced by magnetic field inhomogeneities (referred to as $T_2^*$ in NMR~\cite{levitt2008spin}).

\begin{table}[h]
    \centering
    \includegraphics[scale=0.85]{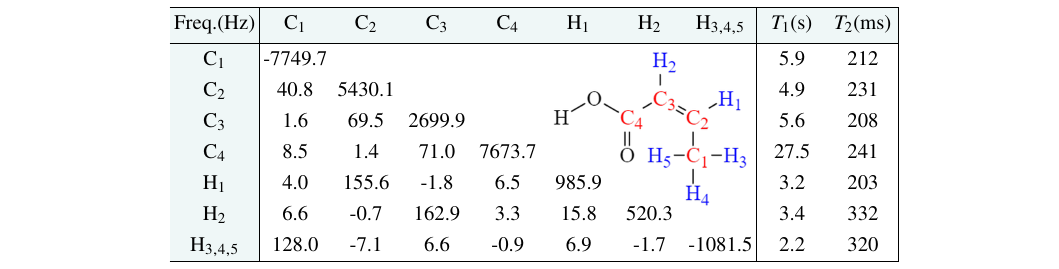}
    \caption{
        Sample parameters and molecular structure of the nine-spin crotonic acid sample. 
        The left portion presents chemical shifts (diagonal elements) and $J$-couplings (off-diagonal elements), while the right lists the corresponding relaxation times.
        The chemical shifts are given in the rotating frame, referenced to 100.6273~MHz for $^{13}$C and 400.2118~MHz for $^{1}$H.}
    \label{tab:sample_param}
\end{table}

The preprocessed input component is encoded in the pulse area of a radiofrequency (RF) field, with its phase aligned along the $x$-axis and frequency resonant with designated nuclei (in this work we consider inputs with ${\rm dim}\leq2$, while higher‑dimensional inputs can be encoded using general parameterized circuits).
Each pulse, with a duration of approximately $10^{-5}$ s, is short enough that the system's free evolution during the pulse can be neglected, thereby effectively implementing a global rotation on all proton spins.
Although slight deviations in the rotation angle and axis may arise from detuning between the chemical shift and the RF frequency, this effect remains fixed and is incorporated within the quantum reservoir. 
The pulses are applied at regular intervals $\tau$, during which the system evolves under its natural Hamiltonian and relaxation dynamics without external control.
The readout is performed at the end of the evolution period to measure the protons, since the signal-to-noise ratio for carbon is significantly lower.
Consequently, the carbon spins are treated as an inaccessible bath, demonstrating that the computational power of the system can be harnessed solely by accessing a subset of its spins.
Two types of readout schemes are implemented: one scheme involves measuring the observable expectations $\langle \sigma^{m}_{x,y,z} \rangle$ with $m={\rm H_1}, {\rm H_2}$, and the methyl protons, while the time multiplexing readout scheme is described in the subsequent section. 
The readout vector dataset is normalized using Z-score standardization followed by ridge regression, with the regularization strength determined via 10-fold cross-validation.

In our experiments, the number of washout steps $n_{\rm wo}$ was determined by extending the washout sequence until the NMR spectral readout (see FIG.~\ref{fig_sm:fid_spec}(b)) remained invariant across different initial states.
For NARMA tasks, we used $n_{\rm wo} = 100$ for $\tau = \SI{0.3}{s}$ and 1000 for $\tau = \SI{0.01}{s}$, which may be larger than the minimum required to achieve convergence.
As described in the main text, a rewinding protocol was employed in our experiments \cite{mujal2023time}; that is, each experiment is reinitialized with last $n_{\rm wo}$ input-evolution cycles (from $k-n_{\rm wo}$ to $k-1$) to reprepare the reservoir to $\rho_{k-1}$, then perform the input-evolution-readout procedure for the $k$-th step.
Other experimental protocols have been reported in recent years \cite{mujal2023time, kobayashi2024feedback}, such as the weak measurement protocol that enables online data processing without restarting experiments, which we plan to explore in future studies.

\setcounter{secnumdepth}{2}
\section{Theoretical basis of time-multiplexed readout}

Time-multiplexed readout provides an efficient way to extract information from a quantum system, particularly when direct access to many observables is not available.
When performed during input-independent evolution, this process is conceptually analogous to \textit{dynamical quantum tomography}, where information about the (input) state is inferred from a time series of measurements of fixed observables~\cite{kech2016dynamical, xiao2024quantum}.
By accessing Heisenberg-evolved observables at multiple evolution times, the time-multiplexed scheme effectively increases the number of independent readout variables, thereby raising the upper bound on the achievable performance.

Formally, consider a $d$-dimensional quantum system initialized in state $\rho$, whose dynamics are described by a quantum channel $\Phi_t(\rho)=\sum_a K_{t,a}\rho K_{t,a}^\dagger$.
The expectation value of an observable $O$ at time $t$ is
\begin{equation}
    \langle O \rangle_t = {\rm Tr}[\rho(t) O]
    = {\rm Tr}\,\Big[\sum_i K_{t,a} \rho K_{t,a}^\dagger O\Big]
    = {\rm Tr}\,\Big[\rho \sum_i K_{t,a}^\dagger O K_{t,a}\Big]
    = {\rm Tr}[\rho\, O(t)],
\end{equation}
where $O(t)=\Phi_t^\dagger(O)$ denotes its Heisenberg-picture evolution. 
The accessible measurement space is thus $\mathcal{M} = {\rm span}\{O(t)\,|\,O\in\mathcal{S},\,t\in\mathcal{T}\}$, with $\mathcal{S}$ the set of directly measurable independent observables and $\mathcal{T}$ the sampling times. 
We use ${\rm dim}(\mathcal{M})$ to denote the dimension of the operator space $\mathcal M$, and $|\mathcal{S}|$ to denote the number of observables in the set $\mathcal S$. 
When the dynamics generate nontrivial operator evolution with $[K_{t, a}, O]\neq 0$, the condition ${\rm dim}(\mathcal{M}) > |\mathcal{S}|$ typically holds, indicating that time multiplexing provides access to a larger set of independent operator components than direct measurement alone.
Each time-evolved operator $O(t)_i$ (where $i$ indexes distinct observable–time pairs) can be expanded in an operator basis as $O(t)_i = \sum_{k=1}^{{\rm dim}(\mathcal{M})} \alpha_{ik} E_k$, yielding the QRC output 
\begin{equation} 
    y = \sum_{i} w_{i} {\rm Tr}[\rho O(t)_i] + b = \sum_{k=1}^{{\rm dim}(\mathcal{M})} w_k'\, {\rm Tr}[\rho E_k] + b,
\end{equation}
which allows the QRC to exploit these additional degrees of freedom.

In the ideal case with sufficient sampling points, ${\rm dim}(\mathcal{M})$ can reach its maximal value, given by the dimension of the Krylov subspace spanned by the Heisenberg-evolved observables $\{\Phi_t^\dagger(\tilde O_i)\,|\, t\geq 0\}$.
The Krylov dimension therefore sets an upper bound on the number of independent readout variables accessible via time multiplexing, and thus limits the reservoir's total information processing capacity (IPC), which quantifies its ability to implement linear memory and nonlinear mappings of time-lagged inputs~\cite{dambre2012information}.
By increasing the number of independent measurements from $|\mathcal{S}|$ to ${\rm dim}(\mathcal{M})$, the time-multiplexed scheme effectively raises the upper bound on the total IPC and enhances the achievable performance of QRC.

When the generators of dynamics are known, ${\rm dim}(\mathcal{M})$ can be inferred analytically.
Here we take the NMR setting used in this work as an example.
Under the internal Hamiltonian $H=\sum_{i=1}^{N}\pi\nu_i\,\sigma_z^i+\sum_{i<j}\frac{\pi}{2}J_{ij}\sigma_z^i\sigma_z^j$, the transverse operator on the first spin $\sigma_y^1+i\sigma_x^1$ evolves in the Heisenberg picture as
\begin{equation}
    e^{iHt} (\sigma_y^1+i\sigma_x^1) e^{-iHt} \propto \sum_{q_n\in\{0,1\}} |1q_2\ldots q_N\rangle\langle 0q_2\ldots q_N| \,\exp{-i\left[2\pi\nu_1 + \pi\sum_{n=2}^N (-1)^{q_n} J_{1n}\right] t},
\label{eq:op_evolve}
\end{equation}
which contains $2^{N-1}$ coherences with distinct oscillation frequencies and can be expanded into $2^N$ Pauli operators corresponding to multi-spin correlations. 
Such information is directly captured in the free-induction-decay (FID) signal, which we introduce in Section~\ref{sec:fid}.
More generally, even without exact knowledge of the dynamics, it has been proven that under sufficiently generic (symmetry-free) evolution, a single observable sampled over time can span the full operator space~\cite{rall2025quantum}, ensuring the effectiveness of the time-multiplexed scheme.

We finally analyze the experimental runtime of time-multiplexed readout in comparison with conventional single-time readout, i.e., measuring multiple specified observables of a static state.
\begin{itemize}
    \item [(a)] 
    On platforms supporting continuous weak measurement, all sampling points $\{O(t_i)\}$ can be obtained within a single experimental run, as the system–detector coupling is sufficiently weak to suppress measurement back-action. 
    In this way, time-multiplexed readout extracts far more effective observables than a single-time measurement of $O$, enabling markedly improved performance without adding extra runtime. 
    Such continuous monitoring schemes are available on various platforms \cite{smith2006efficient, murch2013observing, colangelo2017simultaneous, cujia2019tracking}, including NMR systems, where the standard FID detection naturally provides a weak-measurement time trace.

    \item [(b)]
    On platforms limited to projective measurements, obtaining $M$ sampling points requires $M$ separate experiments. 
    For a fair comparison, single-time readout would also need approximately $M$ experiments to access a comparable number of independent observables, since QRC performance depends strongly on readout dimension \cite{dambre2012information}.  
    Thus, the total experimental runtime of the two approaches is expected to be similar when matched to the same performance target. 
    Parallel measurement of commuting observables does not change this conclusion, because the same strategy can be applied to each experiment in time-multiplexed readout.
    Moreover, time-multiplexed readout is often experimentally favorable, as it uses free evolution and avoids the need to implement dedicated control sequences for different observables.
\end{itemize}

\setcounter{secnumdepth}{3}
\section{Experimental implementation of time-multiplexed readout using FID signal}
\label{sec:fid}

\begin{figure}[htbp]
    \centering
    \includegraphics[scale=0.95]{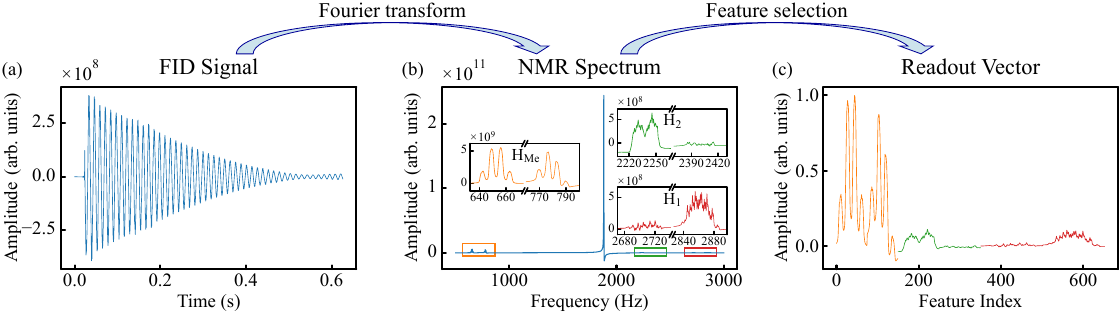}
    \caption{
        Feature extraction from the NMR free-induction-decay (FID) signal.
        {\bf (a)} The real part of the FID signal obtained from the proton spins of crotonic acid, showing the first 2048 points (one quarter of the total acquisition). 
        The signal amplitude is enhanced by the spectrometer's receiver gain.
        {\bf (b)} The real part of the corresponding NMR frequency spectrum, produced by Fourier transforming the FID signal. 
        The intense peak near 1880 Hz originates from $\rm H_2O$ in the solvent. 
        The red, green, and orange boxes highlight the spectral regions of crotonic acid's $\rm H_1$, $\rm H_2$, and methyl protons $\rm H_{Me}$ (i.e., the equivalent $\rm H_3, H_4, H_5$), with insets showing magnified views. 
        Although these signal intensities are two to three orders of magnitude lower than that of the $\rm H_2O$ peak, they still exhibit sufficiently high signal-to-noise ratios.
        {\bf (c)} Normalized readout vector constructed by concatenating the spectral amplitudes from the highlighted regions, serving as time-multiplexed readout features for QRC.
        }
    \label{fig_sm:fid_spec}
\end{figure}

Now we introduce the natural realization of time-multiplexed readout in the NMR platform: the FID signal \cite{levitt2008spin}.
The nuclear spins to be measured undergo free evolution, precessing about the static $z$-axis magnetic field while relaxing toward thermal equilibrium. 
The resulting transverse magnetization induces a proportional oscillating decaying current in a detection coil in the $xy$-plane, thereby generating the FID signal (FIG.~\ref{fig_sm:fid_spec}(a)).
NMR employs quadrature detection, simultaneously measuring two orthogonal components of the complex nuclear magnetic signal.
For the $m$-th observed nucleus, the measurement operator is expressed as $\sigma_y^{m}+i\sigma_x^{m} \propto \dyad{1}{0}^{m}$.
In this work, all proton spins in the crotonic acid sample are detected, and the FID signal is well approximated by $S(t)\propto {\rm Tr}\big[ e^{-iHt} U \rho U^{\dagger} e^{iHt} \sum_{m=5}^{9}\big(\sigma^{m}_{y}+i\sigma^{m}_{x}\big) \big]e^{-\lambda t}$, where $U=\prod_{m=5}^9 R^m_x(\frac{\pi}{2})$ rotates the longitudinal magnetization into the $xy$-plane.
The acquired FID signal contains 8192 complex-valued points sampled at $\SI{0.3}{ms}$ intervals, as shown in Fig.~\ref{fig_sm:fid_spec}(a).

In practice, we do not use the raw FID signal directly, but instead perform a Fourier transform and select the sparsely distributed spectral peaks as readout features, as shown in Fig.~\ref{fig_sm:fid_spec}(b).
This choice is motivated by two considerations. 
First, the time-domain FID contains a large background contribution from $\mathrm{H_2O}$ molecules in the solvent, which can be removed by spectral filtering in the frequency domain, since it appears as a well-isolated peak near 1880~Hz.
Second, discarding frequency components that do not carry reservoir-related signals reduces the effective feature dimension, thereby lowering computational overhead and mitigating potential overfitting.
For the selection of specific features, we take the real spectral amplitudes around the signal peak positions and concatenate them into a 653-dimensional readout vector, as shown in Fig.~\ref{fig_sm:fid_spec}(c).
This includes all major peaks while keeping the feature number low for stable regression.
Small variations in the selected spectral range have a negligible influence on QRC performance.
Further dimensionality reduction is possible because adjacent points within each spectral peak are highly correlated. However, this involves a trade-off between simplicity and predictive accuracy, which can be task-dependent and requires further analysis in future work.

\setcounter{secnumdepth}{4}
\section{The NARMA task setup and experimental prediction results}

The NARMA task setup follows the pioneering work on quantum reservoir computing \cite{fujii2017harnessing}, which has also been adopted in several subsequent studies \cite{suzuki2022natural, yasuda2023quantum, monzani2024leveraging}. 
Two types of NARMA dynamics are considered:
\begin{equation}
\begin{aligned}
    &\text{NARMA2: } y_{k+1}=0.4y_{k}+0.4y_{k}y_{k-1}+0.6s_{k}^3+0.1,\\
    &\text{NARMA$n$: } y_{k+1}=0.3 y_k+ 0.05 y_k\left(\sum_{i=0}^{n-1} y_{k-i}\right)+ 1.5 s_{k-n+1} s_k+ 0.1,
\end{aligned}
\end{equation}
where the order $n$ controls the dependence on long time lags.
The input sequence is constructed using a superimposed sine wave given by
\begin{equation}
    s_k=0.1\left[\sin \left(\frac{2 \pi \bar{\alpha} k}{T}\right) \sin \left(\frac{2 \pi \bar{\beta} k}{T}\right) \sin \left(\frac{2 \pi \bar{\gamma} k}{T}\right)+1\right],
\end{equation}
where $(\bar{\alpha}, \bar{\beta}, \bar{\gamma}, T)=(2.11, 3.73, 4.11, 100)$. 

\begin{figure}[htbp]
    \centering
    \includegraphics[scale=1]{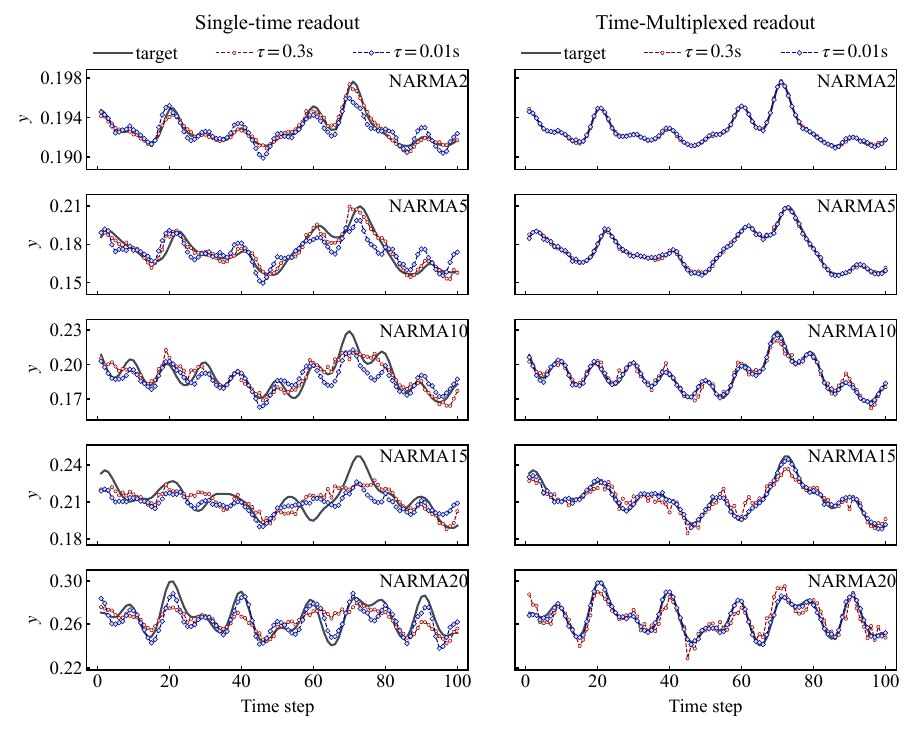}
    \caption{
        Experimental prediction results for NARMA tasks with $n=2,5,10,15,20$, utilizing quantum reservoirs with different evolution durations and readout schemes.
        }
    \label{fig_sm:narma_pred}
\end{figure}

FIG.~\ref{fig_sm:narma_pred} presents the experimental prediction results for NARMA tasks with order $n=2,5,10,15$ and 20, utilizing quantum reservoirs with different configurations.
The left and right panels show results with single-time readout ($\langle \sigma^{m}_{x,y,z} \rangle$) and time-multiplexed readout, respectively, while red and blue lines represent evolution durations of $\tau = \SI{0.3}{s}$ and $\SI{0.01}{s}$, respectively.

\setcounter{secnumdepth}{5}
\section{Numerical characterization of spin-based quantum reservoir dynamics}
\label{sec:num_sim}

In this section, we present the numerical implementation of the input injection and the subsequent system evolution in the quantum reservoir. 
The normalized input $\bar{s}_k \in [0,1]$ is encoded as a rotation angle $\theta_k = \arcsin(\bar{s}_k)$ of an RF pulse applied to a specific nuclear channel (e.g., proton), effecting a global $x$-rotation on the corresponding spin set $S_{\rm in}$:
\begin{equation}
    \prod_{n\in S_{\rm in}} R^{n}_x(\theta_k)
    = \exp\left[-i\, t_k \sum_{n\in S_{\rm in}} \frac{\omega_d}{2}\sigma_x^n \right],
\end{equation}
with a fixed driving frequency $\omega_d$ and an adjusted pulse duration $t_k$ satisfying $\omega_d t_k = \theta_k$.
More precisely, when the internal Hamiltonian evolution during the pulse duration is taken into account, the actual unitary operation becomes
\begin{equation}
    U_{s_k} = \exp\!\left[-i\, t_k \left(\sum_{n\in S_{\rm in}} \frac{\omega_d}{2}\sigma_x^n + H \right)\right],
    \label{eq:input}
\end{equation}
where the system Hamiltonian is
\begin{equation}
    H = \sum_{i=1}^{N} \pi \nu_i\, \sigma^{i}_{z}
      + \sum_{1 \le i < j \le N} \frac{\pi}{2} J_{ij}\, \sigma^{i}_{z} \sigma^{j}_{z}.
\end{equation}
This introduces a slight deviation of the effective rotation axis from the $x$ direction.

The system evolution is governed by the Lindblad master equation:
\begin{equation}
    \dv{\rho}{t} = -i \left[H, \rho\right] + \mathcal{R}[\rho],
\end{equation}
where $\mathcal{R}$ denotes the relaxation channel.
For a single spin, the relaxation channel is given by
\begin{equation}
    \mathcal{R}[\rho] = \frac{1}{2} \sum_{a=+,-,z} \left( 2 L_a \rho L_a^{\dagger} - L_a^{\dagger} L_a \rho - \rho L_a^{\dagger} L_a \right),
\end{equation}
with the collapse operators defined as
\begin{equation}
    L_{+} = \sqrt{\frac{1-p}{2T_1}}\, \sigma_{+}, ~
    L_{-} = \sqrt{\frac{1+p}{2T_1}}\, \sigma_{-}, ~
    L_{z} = \sqrt{\frac{1}{2T_{\phi}}}\, \sigma_{z}.
\end{equation}
Here, $\sigma_{+} = \dyad{1}{0}$ and $\sigma_{-} = \dyad{0}{1}$ denote the spin raising and lowering operators, respectively. 
The operators $L_{+}$ and $L_{-}$ define a generalized amplitude damping channel that drives the spin toward its equilibrium state,
\begin{equation}
    \rho_{\rm eq} = \frac{1}{2} 
    \begin{bmatrix}
        1+p & 0 \\
        0   & 1-p
    \end{bmatrix},
\end{equation}
at a rate $T_1^{-1}$, where $p$ represents the spin polarization (with $p=1$ recovering the standard amplitude damping channel) \cite{nielsen2001quantum}. 
The operator $L_z$ accounts for pure dephasing due to spin-spin interactions and field inhomogeneities, with a characteristic time $T_{\phi}$.
Combined with amplitude damping, the effective decoherence rate is given by
\begin{equation}
    T_2^{-1} = \left(2T_1\right)^{-1} + T_{\phi}^{-1}.
\end{equation}
Consequently, the transverse spin polarization decays exponentially as $e^{-t/T_2}$ while precessing in the $xy$-plane, whereas the longitudinal polarization along the $z$-axis relaxes toward its equilibrium value according to $p_z(t) = p + \left[p_z(0) - p\right] e^{-t/T_1}$.

For an $N$-spin system, the relaxation dynamics become more complex. In our simulations, we assume that the spins relax independently so that the total relaxation channel is given by
\begin{equation}
    \mathcal{R}[\rho] = \sum_{n=1}^{N} \mathcal{R}^{(n)}[\rho],
\end{equation}
with each $\mathcal{R}^{(n)}[\rho]$ characterized by its individual relaxation parameters $T^{(n)}_1$ and $T^{(n)}_2$.

\setcounter{secnumdepth}{6}
\section{Numerical simulation of 9-spin reservoir computing for NARMA tasks}
\label{sec:sim_result}

To validate the reliability of our experimental results, we performed numerical simulations of the nine-spin quantum reservoir computing model on the NARMA task, following the procedure described in Section~\ref{sec:num_sim} and using the parameters listed in Table~\ref{tab:sample_param}. 
The simulations were implemented using the Python package \texttt{QuTiP} \cite{lambert2024qutip}.
For the polarization setting, because the realistic polarization values in room‑temperature NMR (on the order of $10^{-5}$) can lead to numerical instability when solving the Lindblad equation, we scale the carbon and proton polarizations to $[1,\, 3.98] \times 10^{-4}$ (proportional to their gyromagnetic ratios) to balance numerical robustness with experimental realism.

\begin{figure}[htbp]
    \centering
    \includegraphics[scale=1.0]{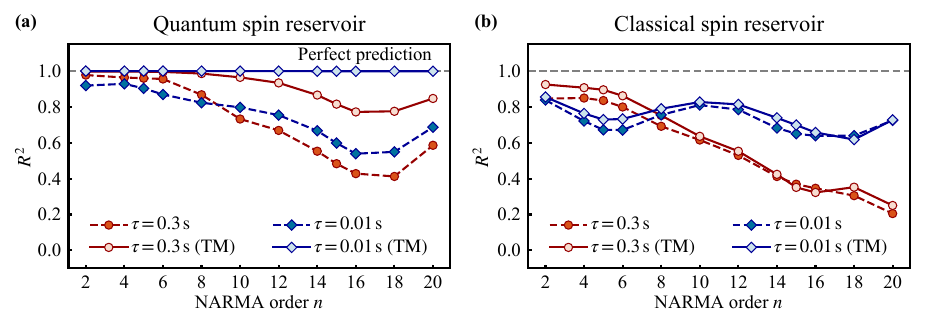}
    \caption{
        Prediction performance (evaluated by $R^2$) on NARMA tasks with $n$ varying from 2 to 20, obtained from numerical simulations of (a) quantum spin reservoir and (b) classical spin reservoir.
        Red and blue lines represent input intervals of $\tau = \SI{0.3}{s}$ and $\SI{0.01}{s}$, respectively, while solid and dashed lines indicate results obtained with time multiplexing and single-time measurements.}
    \label{fig_sm:sim_narma}
\end{figure}

The simulation results, presented in Fig.~\ref{fig_sm:sim_narma}(a), are qualitatively consistent with the experimental results in Fig.~3 of the main text.
Quantitative differences are attributed to deviations between the simulated dynamics and the actual experimental processes, including cross-correlated relaxation \cite{kowalewski2017nuclear} and unidentified systematic errors.
A detailed analysis of experimental imperfections and their effects on QRC performance is provided in Section~\ref{sec:imperfection}. 
Accurate modeling of open quantum many-body dynamics remains a significant challenge and warrants further investigation.

To elucidate the effect of quantumness, we compare our QRC model with the corresponding classical spin reservoir computing model, obtained by treating the spins as classical unit vectors \cite{kornjavca2024large}.
The dynamics of the $i$-th spin are described by 
\begin{equation}
    \begin{aligned}
        \frac{{\rm d}S_{x}^{(i)}}{{\rm d}t} &= -\Omega_{i} S_{y}^{(i)} - \frac{S_{x}^{(i)}}{T_2^{(i)}} \\ 
        \frac{{\rm d}S_{y}^{(i)}}{{\rm d}t} &= \Omega_{i} S_{x}^{(i)} - \frac{S_{y}^{(i)}}{T_2^{(i)}} \\
        \frac{{\rm d}S_{z}^{(i)}}{{\rm d}t} &= \frac{1 - S_{z}^{(i)}}{T_1^{(i)}} 
    \end{aligned},
\end{equation}
where $\Omega_i=2\pi \nu_i + 2\pi\sum_{j\neq i} J_{ij} S_{z}^{(j)}$ represents the effective instantaneous field exerted on spin-$i$ by the surrounding spins.
Hence, the dynamics of $N=9$ classical spins are characterized by $3N=27$ differential equations.
The simulation results for the classical spin reservoir, with experimental settings identical to those used for QRC, are shown in FIG.~\ref{fig_sm:sim_narma}(b).
These results demonstrate that the optimal performance achievable by the classical spin reservoir remains inferior to that of the quantum reservoir, even when time multiplexing is employed.
This is because classical spins lack the additional degrees of freedom, such as quantum correlations, that can be exploited via time multiplexing, thereby highlighting the advantage of quantumness.

\setcounter{secnumdepth}{7}
\section{Robustness analysis against experimental imperfections}
\label{sec:imperfection}

In this section, we analyze the robustness of QRC against different experimental imperfections.
The impact of such imperfections mainly depends on whether they are time-invariant or time-varying.
\begin{itemize}
    \item[(a)] Time-invariant systematic deviations can be regarded as part of the intrinsic reservoir dynamics, effectively transforming the ideal quantum map into a modified yet still deterministic one, whose influence is not necessarily positive or negative.
    In particular, constant linear offsets in the readout can be fully absorbed by linear regression during training, yielding the same prediction accuracy as in the ideal case.
    
    \item[(b)] Time-varying or random errors disrupt the deterministic nature of the reservoir mapping, causing identical input sequences to evolve into inconsistent trajectories over time.
    This randomness blurs the state-space representation, reduces the fidelity of information retrieval, and thus weakens the reservoir's separability, memory capacity, and predictive accuracy.
\end{itemize}
To quantify these effects, we numerically examined three representative imperfections: (a) pulse errors, (b) relaxation-time drift, and (c) measurement noise. 
To reduce computational cost, simulations were performed on a six-spin subsystem ($\mathrm{C}_1$–$\mathrm{C}_3$ and $\mathrm{H}_1$–$\mathrm{H}_3$) of the nine-spin molecule, with all other parameters identical to those used in experiments.
The input interval was fixed at $\tau = 0.01\,\mathrm{s}$, and time-multiplexed readout was employed.
The performance was evaluated using the average prediction accuracy $\overline{R^2}$ across NARMA2–NARMA20 tasks.

\begin{figure}[htbp]
    \centering
    \includegraphics[scale=1]{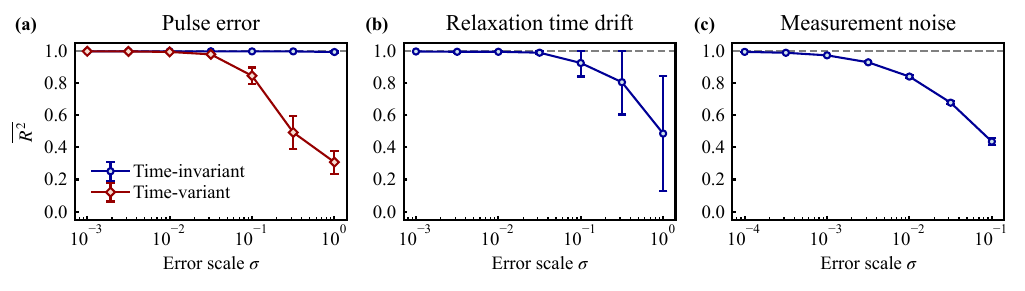}
    \caption{
        Average NARMA performance (evaluated by $\overline{R^2}$) as a function of error scale of (a) pulse strength, (b) relaxation time drift, and (c) measurement noise.
        Simulations use a six-spin subsystem ($\mathrm{C}_1$–$\mathrm{C}_3$ and $\mathrm{H}_1$–$\mathrm{H}_3$) of the nine-spin molecule, with input interval $\tau=0.01\,\mathrm{s}$ and time-multiplexed readout. 
        Each point represents the average over 20 random realizations, with the standard deviation shown as the error bar.
        The dashed line represents the ideal case.
    }
    \label{fig_sm:error}
\end{figure}

\textbf{Pulse imperfections.} 
One relevant source is the finite-duration effect of RF pulses: during the pulse, the system evolves under both the control field and the internal Hamiltonian as described in Eq.~\eqref{eq:input}, resulting in a time-invariant deviation from an ideal rotation.  
In our setting, this induces a slight tilt of the effective rotation axis from the $x$ direction, and changes $\overline{R^2}$ by only 0.001, indicating a negligible impact on overall performance.

Another source is the deviation of RF amplitude, modeled as 
\begin{equation}
    U_{s_k}^{\epsilon} = \exp\!\left[-i\, t_k \left(\sum_{n\in S_{\rm in}} \frac{\omega_d}{2}(1+\epsilon)\sigma_x^n + H \right)\right],
\end{equation}
where $\epsilon \sim \mathcal{N}(0,\sigma^2)$.
Two cases are considered: (i) time-invariant offsets arising from spatial RF inhomogeneity and (ii) time-varying fluctuations representing stochastic drift across different pulses within a single experiment.
As shown in Fig.~\ref{fig_sm:error}(a), time-invariant offsets (blue curve) have almost no influence on $\overline{R^2}$, indicating that such systematic deviations are effectively incorporated into the reservoir dynamics without disrupting its operation, demonstrating the intrinsic robustness of QRC.  
(Note that some static deviations can still degrade performance, such as strong decoherence.)
In contrast, time-varying errors (red curve) lead to a monotonic decline in $\overline{R^2}$ with increasing error scale, decreasing by about 0.004 at $\sigma = 10^{-2}$ and nearly 0.15 at $\sigma = 10^{-1}$.  
In our experiments, the dominant pulse imperfections arise from static RF field inhomogeneity across the sample volume, typically up to 5\% depending on the probe and sample geometry, whereas temporal pulse-to-pulse amplitude fluctuations are generally negligible.

\textbf{Relaxation time drift.} 
Relaxation times may drift over long experimental runs or fluctuate due to environmental variations. 
We model this effect as
\begin{equation}
    T^{\epsilon}_1 = T_{1}^0(1+\epsilon_1), \quad T^{\epsilon}_2 = T_{2}^0(1+\epsilon_2),
\end{equation}
with $\epsilon_i \sim \mathcal{N}(0,\sigma^2)$ updated every 100 steps to emulate slow fluctuations.  
As shown in Fig.~\ref{fig_sm:error}(b), $\overline{R^2}$ decreases by 0.003 at $\sigma = 10^{-2}$ and by 0.073 at $\sigma = 10^{-1}$, indicating strong robustness against small quasi-static drift, while large fluctuations reduce performance due to loss of dynamical determinism (a drop of 0.513 at $\sigma = 10^{0}$).  
In NMR experiments, relaxation time drift may arise from temperature variations, magnetic-field instability, or slow chemical changes in the sample solution.  
In our setup, however, the longest runs last only about 10 hours, during which variations in $T_1$ and $T_2$ are negligible under active temperature stabilization.

Furthermore, if the relaxation times can be accurately tuned, they could serve as an additional resource for improving QRC performance.  
From a dynamical perspective, maintaining $T_1$ and $T_2$ within a comparable range ensures a feasible operating regime.  
As discussed in Section~\ref{sec:T1T2}, the two relaxation processes play competing roles: $T_1$ relaxation drives the system toward the ground or a thermal-equilibrium state with finite polarization, whereas $T_2$ relaxation, together with noncommuting unitary dynamics such as $R_x$, depolarizes the system toward a maximally mixed state with vanishing polarization. 
When $T_2 \ll T_1$, depolarization dominates and the reservoir operates at extremely low polarization, making it practically infeasible because the output signal becomes easily obscured by noise.  
Conversely, tunable relaxation allows for richer noncommuting dynamics while avoiding excessive depolarization, potentially enabling more expressive information processing.  
From a task-adaptation perspective, $T_1$ and $T_2$ determine the reservoir's effective memory depth.  
Thus, tuning them can in principle optimize the match between the reservoir memory and the temporal structure of a given task, thereby improving predictive performance~\cite{sannia2024dissipation}.

\textbf{Measurement noise.}  
Measurement errors are modeled as additive Gaussian noise $\epsilon \sim \mathcal{N}(0,\sigma^2)$ applied to the normalized FID signal.
Each simulated FID signal was normalized by the average maximum amplitude of all FID signals over the dataset, after which Gaussian noise $\epsilon$ was added. 
The parameter $\sigma$ therefore serves as an inverse measure of the signal-to-noise ratio (SNR). 
As shown in Fig.~\ref{fig_sm:error}(c), random noise monotonically degrades performance, with $\overline{R^2}$ decreasing by 0.003, 0.025, 0.157, and 0.563 for $\sigma = 10^{-4}$, $10^{-3}$, $10^{-2}$, and $10^{-1}$, respectively.
When SNR exceeds $10^3$, the impact of measurement noise becomes negligible, consistent with our experimental conditions.

Comparing the three types of imperfections—time-varying pulse errors, relaxation time drift, and measurement noise—we observe that measurement noise is the dominant source of the degradation of QRC performance, as it leads to comparable drops in $\overline{R^2}$ at smaller error scales $\sigma$.
This may arise because dynamical imperfections are partially mitigated by the reservoir's fading memory mechanism, whereas measurement noise directly affects the readout and cannot be compensated by subsequent processing. 
These findings provide practical guidance for improving experimental implementations of QRC.  
In addition, we suggest that system stability can be promoted not only by improving experimental hardware but also by leveraging the native operations and intrinsic dynamics of the physical platform, enabling simpler, faster, and more reliable realizations of QRC.

\setcounter{secnumdepth}{8}
\section{Weather forecasting and echo state networks}

\begin{figure}[htbp]
    \centering
    \includegraphics[scale=1]{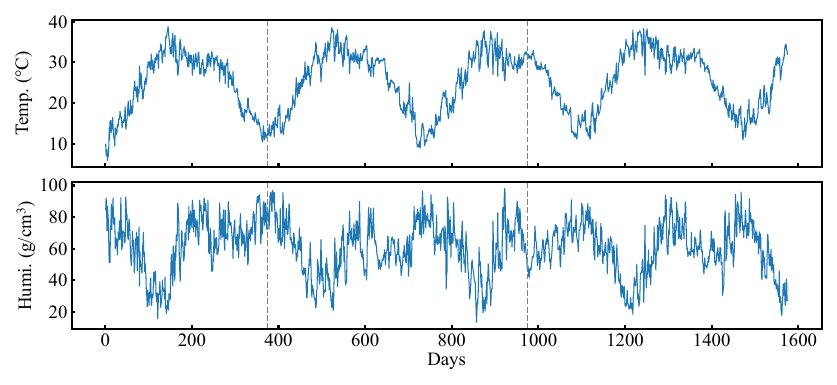}
    \caption{
        Temperature and humidity trajectories from the daily weather dataset. The vertical dashed lines indicate the boundaries between the washout, training, and testing sets.}
    \label{fig_sm:weather_data}
\end{figure}

The weather dataset referenced in the main text is sourced from Kaggle \cite{daily-climate}, a prominent online platform for data science competitions and resources. 
The dataset provides daily weather data for Delhi, India, covering 1,575 days from January 1, 2013, to April 24, 2017. 
It includes four features: temperature, humidity, wind speed, and air pressure. 
For simplicity of experiments, this study focuses on the first two features, temperature and humidity, whose values are presented in FIG.~\ref{fig_sm:weather_data}.
Direct multi-step forecast is performed for the two-dimensional temporal sequence by training separate reservoir weights to model the mapping from $\{{\bf y}_k\}$ to $\{{\bf y}_{k+h}\}$ for different forecast horizons $h$, effectively enabling multitasking \cite{taieb2012review}.
The dataset is partitioned into 374 steps for washout, 600 steps for training, and the remaining steps for testing. 
For the testing phase, 600 steps are used for one-step-ahead prediction, as shown in Fig.~4(a) of the main text, while 556 steps are used for multi-step-ahead prediction with a maximum forecast horizon of $h_{\max}=45$.

\begin{figure}[htbp]
    \centering
    \includegraphics[scale=1]{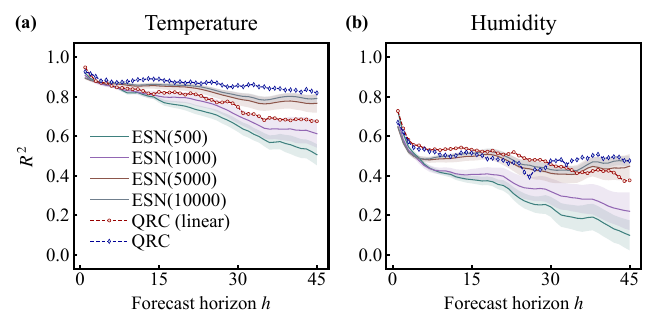}
    \caption{
        Multi-step-ahead forecasting performance results obtained using support vector regression with a radial basis function kernel as a nonlinear processing method on the reservoir readouts.
        }
    \label{fig_sm:svr}
\end{figure}

The baseline model employs Echo State Networks (ESNs), a widely used reservoir computing model based on recurrent neural networks \cite{jaeger2004harnessing}. 
For an ESN with $M$ network nodes (denoted as ${\rm ESN}(M)$), the state update and output are described by
\begin{equation}
    {\bf x}_{k} = \tanh\left( {\bf W}_{\rm in} \begin{bmatrix} 1 \\ {\bf s}_{k} \end{bmatrix} + {\bf W} {\bf x}_{k-1} \right),~~y_k =  \begin{bmatrix} 1 & {\bf x}^{\top}_{k} \end{bmatrix} {\bf w}_{\rm out},
\end{equation}
where ${\bf W}_{\rm in} \in \mathbb{R}^{M \times (1 + d_{\rm in})}$, ${\bf W} \in \mathbb{R}^{M \times M}$ and ${\bf w}_{\rm out}$ are the input, recurrent, and output weights, respectively. 
Only the output weights ${\bf w}_{\rm out}$ are trained to minimize the prediction error, while ${\bf W}_{\rm in}$ and ${\bf W}$ are randomly initialized beforehand and remain fixed during training.

\begin{figure}[htbp]
    \centering
    \includegraphics[scale=0.915]{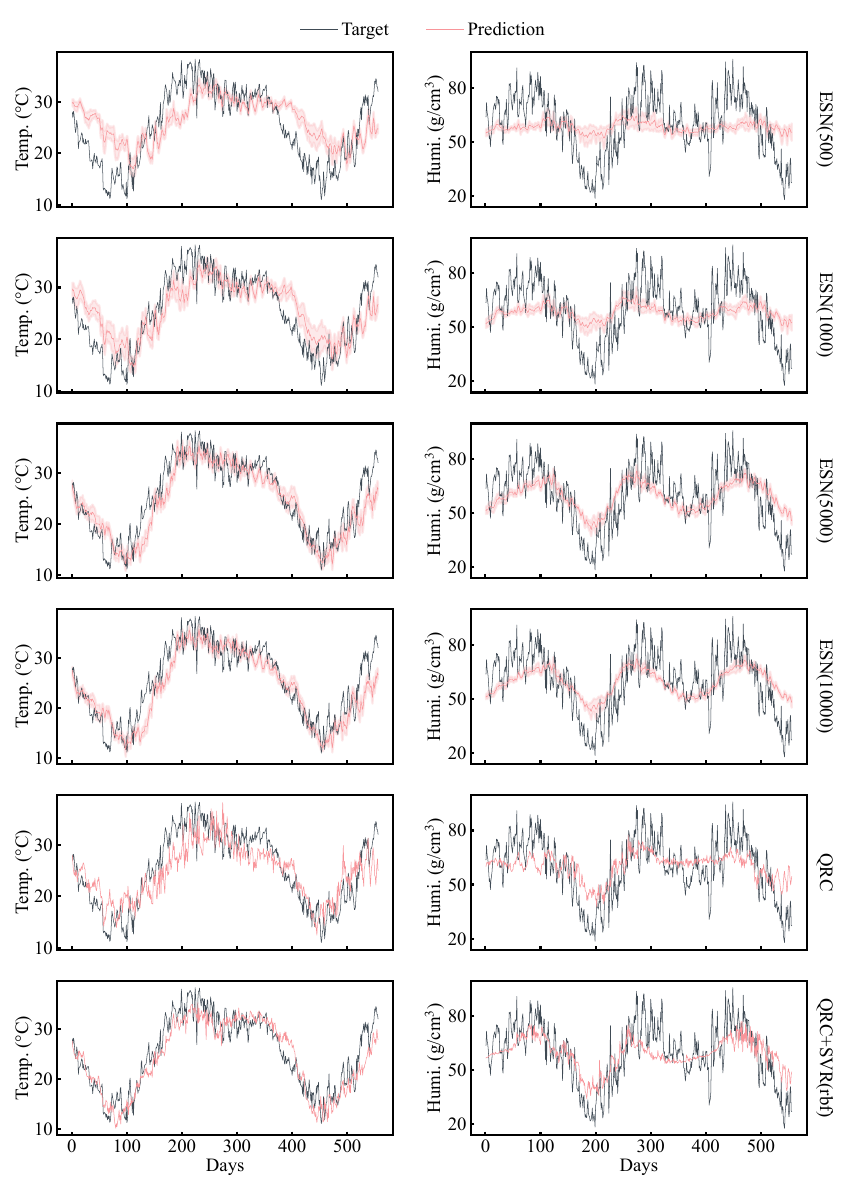}
    \caption{
        45-day-ahead predictions of temperature and humidity. 
        The blue and orange curves represent the target and predicted trajectories, respectively. 
        From top to bottom, the models used are: ESN(500), ESN(1000), ESN(5000), ESN(10000), QRC with linear regression, and QRC with support vector regression using a nonlinear radial basis function kernel.
        The predictions of ESNs are averaged over 100 realizations, with shaded regions representing the standard deviation.
        }
    \label{fig_sm:weather_pred}
\end{figure}

We implemented the ESNs using the Python package \texttt{reservoirpy} \cite{trouvain2020reservoirpy}, where ${\bf W}_{\rm in}$ is sampled from a Bernoulli distribution over $\{-1, 1\}$ and ${\bf W}$ is sampled from a standard normal distribution. 
The connectivity $k$ and spectral radius $r$ of the recurrent weight matrix ${\bf W}$, defined as the fraction of nonzero entries and the largest absolute eigenvalue, respectively, are closely related to the computational power of ESNs. 
In this work, these hyperparameters were determined via grid search, yielding $(k, r)$ pairs of (0.025, 0.99), (0.01, 0.99), (0.0025, 0.99), and (0.0025, 0.99) for ESN(500), ESN(1000), ESN(5000), and ESN(10000), respectively.
FIG.~\ref{fig_sm:weather_pred} presents the 45-day-ahead prediction results of different models.

We note that for both QRC and ESNs, the prediction accuracy for humidity is generally lower than that for temperature. 
This difference mainly reflects their distinct statistical and physical characteristics \cite{krishnamurthy2019predictability, ruiz2022exploring}. 
Temperature evolution is primarily governed by large-scale, slowly varying processes such as diurnal and seasonal cycles and long-term ocean–atmosphere coupling, resulting in smooth, low-frequency dynamics with strong temporal correlations that facilitate forecasting from past values. 
In contrast, humidity is strongly influenced by rapidly changing local processes, including convection, pressure perturbations, and precipitation, which introduce high-frequency fluctuations and weak autocorrelations, thereby limiting predictability. 
Incorporating additional meteorological variables, e.g., pressure and precipitation fields, would likely enhance humidity prediction and enable more comprehensive weather forecasting.

We evaluated support vector regression (SVR) with a radial basis function (RBF) kernel as a nonlinear post–processor for the reservoir readouts.  
Our implementation employed the SVR module from the \texttt{scikit-learn} library \cite{scikit-learn}, with the regularization parameter $C$ and the tolerance margin $\epsilon$ systematically tuned. 
FIG.~\ref{fig_sm:svr} reports the best results obtained.
We observe that SVR(RBF) markedly improves temperature prediction for QRC but provides weaker or negligible benefits for ESNs and for humidity prediction. 
For the difference between QRC and ESNs, we attribute the effect mainly to the dimensionality of the reservoir readout. 
For QRC (653 features) and ESN(500), the feature dimension remains in a regime where the RBF kernel can provide a meaningful nonlinear mapping and yield observable improvements, with the larger gain of QRC likely related to its quantum-enhanced expressibility. 
As the ESN size increases to 1000, 5000, or 10000 nodes, the higher dimensionality causes the RBF kernel to suffer from the curse of dimensionality, where distances between different data points become nearly indistinguishable \cite{donoho2000high}. 
This limits SVR to extract useful nonlinear structure and increases the risk of overfitting \cite{rahimi2007random, braga2015improving}, diminishing the benefit of enlarging the network and leading to negligible or even negative gains over linear regression. 
In the case of humidity prediction, many relevant factors (e.g., pressure and precipitation) are not present in the dataset, which fundamentally limits its predictability and renders nonlinear post-processing much less effective.
These interpretations are consistent with the observed behavior, while a more complete understanding may require further investigation.

\setcounter{secnumdepth}{9}
\section{Effects of $T_1$ and $T_2$ relaxation on the fading memory in quantum reservoir computing}
\label{sec:T1T2}

In this section, we examine the influence of the $T_1$ and $T_2$ relaxation on the fading memory of the quantum reservoir, as quantified by the short-term memory (STM) capacity, through numerical simulations.
STM capacity quantifies the ability of the reservoir to reconstruct past inputs from its current state, which is a fundamental property for processing temporal information \cite{jaeger2001short}.

First, we reiterate the necessity of incorporating relaxation processes. 
As mentioned in the main text, an effective quantum reservoir should possess a high-dimensional state space, nonlinear input–output transformations, and fading memory. 
In unitary systems (commonly employed in quantum algorithm frameworks) the first two properties emerge naturally: the exponential scaling of the Hilbert space provides a high-dimensional representation, while nonlinearity arises from two mechanisms: (i) the input encoding (e.g., rotations yielding sine and cosine responses) \cite{govia2022nonlinear} and (ii) the multiplicative interaction between sequential inputs \cite{fujii2017harnessing, perez2020data}. 
In contrast, fading memory requires that, after a sufficiently long input sequence, the reservoir state converges to a unique trajectory independent of its initial conditions. 
Due to the distance-preserving nature of unitary dynamics, nonunitary processes (such as $T_1$ and $T_2$ relaxation) are essential for gradually erasing old information.

We then assess how these relaxation processes affect the STM capacity. 
For a random input sequence $\{s_k\}$, the target output for a delay $t_d$ is defined as
\begin{equation}
    y_{k} = s_{k-t_d}.
\end{equation}
The STM capacity for delay $t_d$, denoted by $C(t_d)$, is given by
\begin{equation}
    C(t_d)=\frac{{\rm cov}^2({\bf y},{\bf \hat y})}{\sigma^2({\bf y}) \sigma^2({\bf \hat y})},
\end{equation}
where ${\bf y}$ and ${\bf \hat y}$ represent the target and predicted output sequences, respectively. 
The total STM capacity is then obtained by summing over all delays,
\begin{equation}
    C_{\rm STM}=\sum_{t_d=0}^{t_d^{\max}} C(t_d),
\end{equation}
where $t_d^{\max}$ is chosen to 100 in our simulations such that $C(t_d)$ becomes negligible for larger delays.

\begin{table}[htbp]
\setlength\tabcolsep{0pt}
\renewcommand{\arraystretch}{1.3}
\centering
\begin{tabular}{|p{1.4cm}<{\centering}|p{1.2cm}<{\centering}p{1.2cm}<{\centering}p{1.2cm}<{\centering}|p{1.2cm}<{\centering}p{1.2cm}<{\centering}|}
\hline
 \rowcolor{color1} Freq.(Hz) & C       & H      & F      & $T_1$(s)  & $T_2$(ms)    \\ \hline
\cellcolor{color1} C & 79.8   &        &        & 2.9 & 203 \\
\cellcolor{color1} H & 160.6  & 201.0 &        & 2.8 & 236 \\
\cellcolor{color1} F & -194.3 & 48.0  & 157.2 & 3.1 & 200 \\ \hline
\end{tabular}
\caption{
    Parameters of the three-spin diethyl fluoromalonate sample. 
    The left portion presents chemical shifts (diagonal elements, given in the rotating frame) and $J$-couplings (off-diagonal elements), while the right lists the corresponding relaxation times.
}
\label{tab:3-spin}
\end{table}

In our numerical simulations, we employ a three‑spin system with parameters drawn from diethyl fluoromalonate, a widely used NMR molecular sample, as detailed in Table~\ref{tab:3-spin}.
The spin polarizations are set to $[1,\, 3.98,\, 3.74] \times 10^{-4}$ for carbon, proton, and fluorine respectively.
Input is injected into all three spins via the $x$-axis pulse.
Considering that the input interval affects the memory capacity, evolution duration $\tau$ is varied from {$10^{-4}~{\rm s}$} to {$10^{1}~{\rm s}$} to explore its influence.
To avoid the uncertainty introduced by selecting partial measurements, the readout is obtained by measuring $4^3-1=63$ Pauli operators to capture the full state information. 
These expectation values are then recalibrated by dividing by the polarization and further perturbed by adding Gaussian noise with $\sigma = 10^{-5}$ to improve the robustness of linear regression. 
For the dataset, we generate a prolonged random sequence $\{s_k\}$ uniformly distributed in $[0,1]$. 
The training and test sets are taken from the tail of the sequence, with sizes of 3000 and 1000 points, respectively. 
A washout period of size $\max\left\{\left\lceil 10/\tau \right\rceil, t_d^{\max}\right\}$ is applied before the training data to eliminate the effects of the initial state, lasting approximately 10 seconds.

\begin{figure}[htbp]
    \centering
    \includegraphics[scale=1]{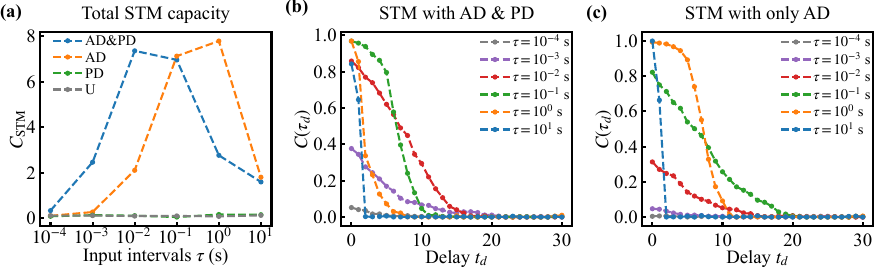}
    \caption{
    {\bf (a)} Total short-term memory (STM) capacity $C_{\rm STM}$ as a function of the input interval $\tau$, comparing four scenarios: amplitude damping only (AD, corresponding to $T_1$ relaxation), phase damping only (PD, corresponding to $T_2$ relaxation), combined amplitude and phase damping (AD \& PD), and unitary dynamics (U). 
    {\bf (b)} and {\bf (c)} depict the STM capacity $C(t_d)$ versus delay $t_d$ for the combined amplitude and phase damping case and for amplitude damping only, respectively.
    }
    \label{fig_sm:STM}
\end{figure}

FIG.~\ref{fig_sm:STM}(a) illustrates the total STM capacity $C_{\rm STM}$ for various configurations of input intervals and relaxation mechanisms. 
As expected, $C_{\rm STM}$ vanishes under unitary evolution due to the absence of fading memory. 
In a unitary system, the reservoir state retains the full influence of the initial state and all prior inputs, preventing the selective extraction of relevant recent information.
However, although the introduction of $T_2$ relaxation (phase damping) leads to some information erasure, it alone is insufficient to achieve meaningful fading memory. 
Under unitary dynamics, the quantum system continuously interconverts population and coherence, while dephasing gradually eliminates coherence. 
This interplay ultimately depolarizes the reservoir, driving it toward the trivial maximally mixed state $\mathds{1}/2^N$. 

Thus, not every dissipative process is suitable for quantum reservoir computing; the key requirement is a process that drives the system toward a nontrivial state—a role fulfilled by $T_1$ relaxation (generalized amplitude damping). 
Several studies have underscored the necessity of non-unital channels (i.e., channels that do not preserve the identity) in quantum reservoir computing \cite{martinez2023quantum, monzani2024leveraging}. 
From an information-theoretic perspective, we conjecture that entropy-decreasing processes are essential for achieving effective fading memory. 
In amplitude damping channels, the system is driven toward an equilibrium state with lower entropy than the maximally mixed state. 
Moreover, previous schemes employing input qubit reset \cite{fujii2017harnessing} and probabilistic state reset \cite{chen2020temporal} incorporate entropy-reducing state preparation processes. 
In contrast, unitary operations preserve entropy, whereas dephasing increases it \cite{nielsen2001quantum}.

The configuration incorporating $T_1$ relaxation exhibits a nonzero STM capacity, with $C_{\rm STM}$ initially increasing and then decreasing as the input interval varies. 
FIG.~\ref{fig_sm:STM}(b) and (c) show the corresponding behavior of $C(t_d)$ as a function of delay. 
In the long-time limit ($\tau \gg T_1$), the reservoir relaxes to an equilibrium state independent of the inputs, causing $C_{\rm STM}$ to vanish. 
For relatively large $\tau$, relaxation erases most of the input history while only retaining the most recent inputs, manifested as a narrow plateau ($C(t_d) \approx 1$) followed by a rapid decline in $C(t_d)$.
As $\tau$ decreases, the reservoir incorporates information from more input steps; however, nonlinear mixing of these inputs can reduce the effectiveness of linear memory retrieval, resulting in an initial $C(0) < 1$ and a gradual decay to zero at larger $t_d$. 
With further decrease in $\tau$, the overall values of $C(t_d)$ diminish, and in the short-$\tau$ limit, the system behaves similarly to a unitary system, lacking effective fading memory.
These results suggest the existence of an optimal input interval that balances the number of retained inputs with high memory fidelity at each delay.

The results incorporating both $T_1$ and $T_2$ relaxation (combined amplitude and phase damping) indicate that even in the presence of strong decoherence, the quantum reservoir remains capable of nontrivial information processing.
In the case of only amplitude damping, the effective dephasing rate is reduced to $(2T_1)^{-1}$.
Consequently, the optimal total STM capacity is reached at a larger $\tau$ compared to the combined $T_1$ and $T_2$ case, as $T_1$ is roughly one order of magnitude larger than $T_2$ (see TABLE~\ref{tab:3-spin}). 
Although one might expect that omitting $T_2$ dephasing would better preserve information encoded in coherence, simulations show no significant improvement in STM capacity. 
This is likely because a slow relaxation ($\sim 10^{-1} ~{\rm Hz}$) tends to take effect over a long evolution ($\sim 10^{0} ~{\rm s}$), during which the high-frequency Hamiltonian ($\sim{10^{2}}~{\rm Hz}$) causes the inputs to be largely mixed in a nonlinear manner.
In this context, the additional dephasing appears beneficial for managing memory.
Overall, improved performance is anticipated by jointly tuning both the relaxation rate and the input interval to suit specific task requirements \cite{sannia2024dissipation}.

\end{document}